\documentclass[12pt,preprint]{aastex}
\usepackage{rotating}
\usepackage{epsfig}

\begin{document}

\title{An Ionized Outflow from AB Aur, a Herbig Ae Star with a Transitional Disk}

\author{Luis F. Rodr{\'\i}guez\altaffilmark{1,2}, Luis A. Zapata\altaffilmark{1},
Sergio A. Dzib\altaffilmark{3}, Gisela N. Ortiz-Le\'on\altaffilmark{1},
Laurent Loinard\altaffilmark{1}, Enrique Mac{\'\i}as\altaffilmark{4}, 
Guillem Anglada\altaffilmark{4}}

\altaffiltext{1}{Centro de Radioastronom\'\i a y Astrof\'\i sica, 
UNAM, Apdo. Postal 3-72 (Xangari), 58089 Morelia, Michoac\'an, M\'exico}
\altaffiltext{2}{Astronomy Department, Faculty of Science, King Abdulaziz University, 
P.O. Box 80203, Jeddah 21589, Saudi Arabia}
\altaffiltext{3}{Max Planck Institut f\"ur Radioastronomie, Auf dem H\"ugel 69, 53121
Bonn, Germany}
\altaffiltext{4}{Instituto de Astrof\'\i sica de Andaluc\'\i a (CSIC), Apartado 3004, E-18080 Granada, Spain}

\email{l.rodriguez@crya.unam.mx}
 
\begin{abstract}
AB Aur is a Herbig Ae star with a transitional disk. Transitional disks present substantial 
dust clearing in their inner regions,
most probably because of the formation of one or more planets, although other explanations are still viable.
In transitional objects, accretion is found to be about an order of magnitude smaller than
in classical full disks. Since accretion is believed to be correlated with outflow activity,
centimeter free-free jets are expected to be present in association with these systems,
at weaker levels than in classical protoplanetary
(full) systems. We present new observations of the centimeter radio emission associated
with the inner regions of AB Aur and conclude that the morphology, orientation, spectral
index and lack of temporal variability of the centimeter source imply the
presence of a collimated, ionized outflow. The radio luminosity of this radio jet is,
however, about 20 times smaller than that expected for a classical system of
similar bolometric luminosity. 
We conclude that centimeter continuum emission is present in association with stars
with transitional disks, but at levels than are becoming detectable only with the upgraded
radio arrays. On the other hand, assuming that the jet velocity is 300 km s$^{-1}$,
we find that the ratio of mass loss rate to accretion rate in AB Aur is $\sim$0.1,
similar to that found for less evolved systems.

\end{abstract}  

\keywords{
ISM: jets and outflows -- 
stars: individual (AB Aur) --
stars: pre-main sequence  --
stars: radio continuum 
}

\section{Introduction}
A significant fraction ($\sim$20\%, Andrews et al. 2011) of the millimeter-bright disk population
shows a dust-depleted cavity around the central star.
These cavities were first indirectly inferred from the
infrared spectra of the star-disk system (e.g. Strom et al. 1989; Calvet et al. 2002;
D'Alessio et al. 2005; Kim et al. 2013) and more recently spatially resolved in images
at infrared (e.g. Geers et al. 2007) and millimeter wavelengths
(e.g. Brown et al. 2009; Andrews et al. 2011; van der Marel et al. 2013;
Perez et al. 2014). These disks are referred to as transitional disks since
the cavity could have been created by dynamical clearing due to tidal interactions with recently formed low-mass companions,
brown dwarfs or giant planets on long-period orbits. However, the origin of the cavity remains controversial,
with viscous accretion,
photoevaporative winds, dust size evolution, and tidal interactions with stellar or planetary companions remaining 
as possible explanations (Williams \& Cieza 2011). In addition, recent studies in the infrared suggest that
some transitional disks may retain a small inner disk, having a gap at intermediate radii instead of
a cavity that goes all the way to the star (Espaillat et al. 2012). These disks with
a gap instead of a cavity  are referred to as pre-transitional disks
and are believed to precede in time the transitional disks. Additional studies are 
needed to clarify the nature of transitional and
pre-transitional disks in order to better understand the formation of planetary systems.  

Stars with transitional disks were expected to show little or no accretion given the apparent lack of
material in the immediate surroundings of the star.
However, the presence of accretion (as inferred from ultraviolet/optical excess emission)
was found in most stars with transitional disks, albeit at rates about an order of magnitude lower
that in classical disks (Najita et al. 2007; Espaillat et al. 2012; Espaillat et al. 2014).  
Several mechanisms have been proposed to explain the apparent paradox of simultaneous millimeter transparency
and accretion (e.g. Rosenfeld et al. 2014).

Since accretion and ejection in protostars and young stellar objects
are correlated, we have started a program to
study stars with transitional or pre-transitional disks in the centimeter radio continuum.
It is known that collimated outflows from these systems are (partially) ionized and
can be detected as weak free-free sources (e.g. Eisl\"offel et al. 2000).
From observations of the [Ne II] 12.81 $\mu$m line,
Pascucci \& Sterzik (2009) have shown that photoevaporative winds with velocities of a few km s$^{-1}$
are present in several transitional disks. The radio free-free is expected to trace a different
component, faster and more collimated than a photoevaporative wind. 

AB Aur is a Herbig Ae star at a distance of 144 pc (van den Ancker et al. 1997; Oppenheimer et al. 2008),
with a mass of 2.4$\pm$0.2 $M_\odot$, a total luminosity of $\sim$38 $L_\odot$ and an estimated age 
of 4$\pm$1 Myr (DeWarf et al. 2003).
AB Aur was considered for many years as the prototype of the Herbig
Ae star surrounded by a large envelope and a disk. However, recent sensitive observations 
have modified this picture. Near-IR observations
from Subaru by Fukagawa et al. (2004) revealed spiral arms of enigmatic origin
in the disk. Using the IRAM array to image the CO(2-1) and $^{13}$CO(2-1) transitions and the continuum 
emission at 1.3mm, Pi\'etu et al.
(2005) found that the dust disk is truncated at an inner radius of about 70 AU 
from the central star. Using the SMA, Lin et al. (2006) found molecular spiral
arms in the CO(3-2) emission. Hashimoto et al. (2011) has set an upper limit of 5 $M_J$ for a possible
companion to AB Aur.

Tang et al. (2012) mapped the CO(2-1) and the 1.3 mm continuum emissions combining data from 
the PdBI array and the SMA, confirming the results of Pi\'etu et al. (2005).
However, these sensitive observations revealed the presence of a CO disk 
inside the cavity with an inclination angle slightly different from that measured in the CO
outer disk. This molecular gas shows a velocity gradient and is consistent with
an inner disk whose rotation axis is at a position angle of $142^{\circ} \pm 1^\circ$.
Rodr\'\i guez et al. (2007) obtained high angular resolution, high-sensitivity Very Large Array observations at 3.6 cm
and confirmed the existence of radio continuum emission associated with the star
(G\"udel et al. 1989; Skinner et al. 1993). Rodr\'\i guez et al. (2007) also detected a new,
faint protuberance that extended 
about $0\rlap.{''}3$ to the SE of AB Aur, at a position angle of $139^\circ \pm 12^\circ$.
Since this protuberance is well aligned with the rotation axis of the inner disk detected in CO,
it could well be a one-sided collimated outflow, as observed in other
sources (i.e. DG Tau; Rodr\'\i guez et al. 2012). The possibility of the radio protuberance tracing
a faint radio companion was also discussed by these authors, without reaching a firm conclusion. 
In this paper we present new, very sensitive radio observations to better understand the
nature of the centimeter emission. 

\section{Observations}

The observations were made with the Karl G. Jansky Very Large Array (VLA) of NRAO\footnote{The National 
Radio Astronomy Observatory is a facility of the National Science Foundation operated
under cooperative agreement by Associated Universities, Inc.} centered at a rest frequency of 8.9 GHz (3.3 cm) during
2012 December.  At that time the array was in its A configuration.  
The phase center was at $\alpha(2000) = 04^h~ 55^m~ 45\rlap.^s8$;
$\delta(2000)$ = $+$30$^\circ~ 33'~ 03.99''$. The absolute amplitude calibrator was J0137$+$3309 and
the phase calibrator was J0443$+$3441. 

The digital correlator of the VLA was configured in 16 spectral windows of 128 MHz width divided 
in 64 channels of spectral resolution. The total bandwidth for the 
continuum observations was about 2.0 GHz in a dual-polarization mode.

The data were analyzed in the standard manner using the CASA (Common Astronomy Software Applications) package of NRAO using
the pipeline provided for VLA\footnote{https://science.nrao.edu/facilities/vla/data-processing/pipeline} observations. 
Maps were made using natural weighting in order to obtain a slightly better sensitivity. 
The resulting image {\it r.m.s} was 2.9 $\mu$Jy beam$^{-1}$ at an angular 
resolution of $0\rlap.{''}28 \times 0\rlap.{''}25$ 
with PA = $+45.1^\circ$.

In Figure 1 we show the 3.3 cm emission of AB Aur overlaid on an image of the 1.3 mm emission
from Tang et al. (2012). The total emission at 3.3 cm is 136$\pm$6 $\mu$Jy and the source is elongated,
with deconvolved dimensions of $0\rlap.{''}17 \pm 0\rlap.{''}02 \times \leq 0\rlap.{''}06 \pm 0\rlap.{''}02;~
PA = 161^\circ \pm 7^\circ$. The major axis of the 3.3 cm emission is approximately
parallel to the rotation axis of the high-velocity CO gas studied by Tang et al. (2012) that
has $PA = 142^\circ \pm 1^\circ$.   

\begin{figure}
\centering
\includegraphics[angle=0,scale=0.8]{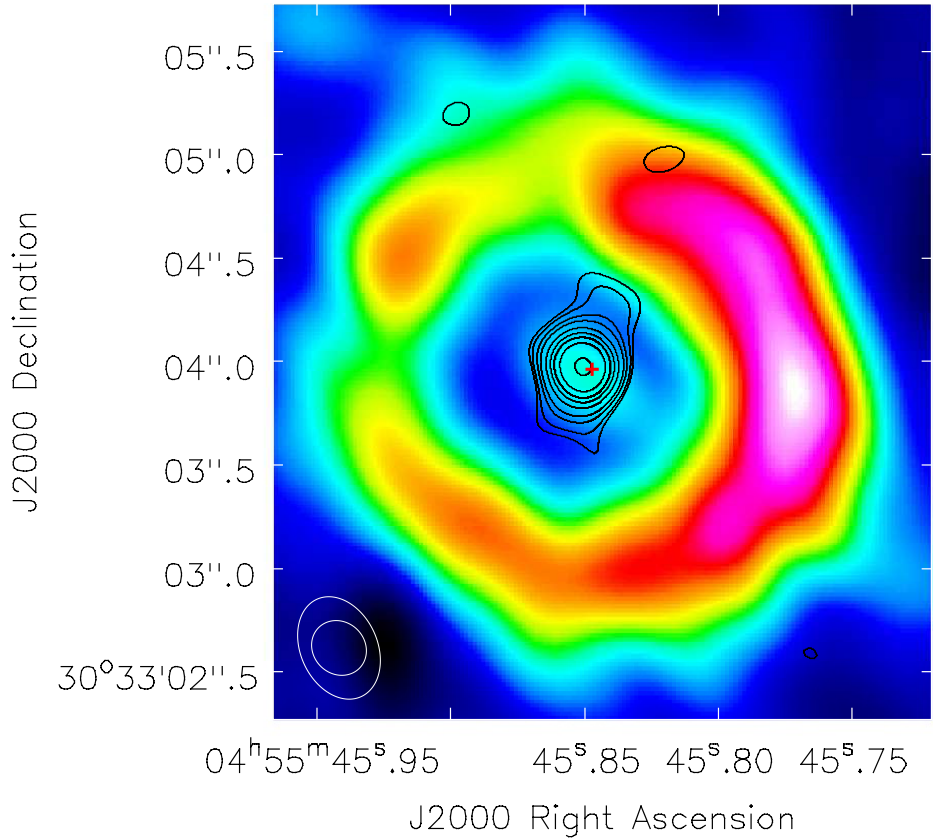}
\vskip-1.0cm
\caption{\small The VLA 3.3 cm continuum 
emission is shown in black contours overlaid on a color image
of the 1.3 mm emission (Tang et al. 2012). 
The black contours are 8.6, 11.5, 17.3, 23.0, 34.6, 46.1, 57.6, 74.9 and 115.2
$\mu$Jy beam$^{-1}$.
The half-power contour of the synthesized beam of the 3.3 cm emission is 
shown as the smaller ellipse in the bottom left corner. The larger ellipse
is the half-power contour of the synthesized beam of the 1.3 mm emission.
The small red cross indicates the position of the star AB Aur, corrected for proper
motions (van Leeuwen 2007).
}
\label{fig1}
\end{figure}

To obtain additional information on the radio spectrum of AB Aur,
Q band ($\sim7$mm) observations from the VLA archive were analyzed.
These observations were taken in the D (observed in three runs during
2010 August 10 and 12 and 2010 September 02) and C (2010 November 19) configurations
(VLA project code: AC982). The flux calibrator was 3C147, while the phase calibrator was J0443+3441. The
phase center for all the observations was R.A.(J2000) $=04^h 55^m 45\rlap.^s944$,
Dec(J2000) $=+30^{\circ}33'05\rlap."79$.
The digital correlator of the JVLA was configured in 16 spectral windows of 128 MHz width divided in 64
channels of spectral resolution. The total bandwidth for the continuum observations was 2 GHz in dual-polarization mode.

Data editing and calibration were performed also using the data reduction package CASA,
following the standard high-frequency JVLA procedures. An averaging in time
($\sim9$ seconds) and in channels (4 channels$=$8 MHz) was performed in order to reduce the volume of the data.

Cleaned images at 7mm were made with the task \textit{clean} of CASA by using the multi-scale multi-frequency
deconvolution algorithm described in Rau \& Cornwell (2011). By concatenating the data from both configurations
(C and D) and using natural weighting, we obtained an r.m.s of 25 ${\mu}Jy$ beam$^{-1}$ and a synthesized beam
of $0\rlap."90\times 0\rlap."70$ with PA$=-65\rlap.^\circ5$. We detect an unresolved source
(Figure 2) centered at the position
R.A.(J2000) $=04^h 55^m 45\rlap.^s846\pm0\rlap.^s002$,
Dec(J2000) $=+30^{\circ}33'04\rlap."04\pm0\rlap."02$. The total flux density at 7mm is 0.84$\pm$0.03 mJy.

\section{Interpretation}

\subsection{The nature of the radio continuum emission}

AB Aur has been detected previously as a centimeter continuum source.
Using the VLA, G\"udel et al. (1989) detected it at 3.6 cm in 1988 October 7 with a flux density of
146$\pm$25 $\mu$Jy and at 3.6 and 6 cm in 1988 October 27 with 
flux densities of 112$\pm$24 $\mu$Jy and 119$\pm$20 $\mu$Jy, respectively.
These authors noted that the radio spectrum deviates from the simple
$\nu^{0.6}$ power law expected for a conical free-free outflow and suggested that the 6 cm emission
could have a non-thermal contribution.
Skinner et al. (1993) detected AB Aur at 3.6 cm with flux densities of
140$\pm$20 $\mu$Jy and 150$\pm$50 $\mu$Jy in data taken on 1990 February
10, 11, and 13 and in 1991 June 1, respectively.
In addition, AB Aur was detected at 3.6 cm with a flux density of 200$\pm$30 $\mu$Jy
in 2006 April 28 (Rodr\'\i guez et al. 2007). 
From observations made in 2011 February, April and May, Dzib et al. (2014, in preparation)
report flux densities for AB Aur of 89$\pm$19 $\mu$Jy (at 6.7 cm = 4.5 GHz) and
167$\pm$25 $\mu$Jy (at 4.0 cm = 7.5 GHz). 
The flux density detected by us at 8.9 GHz (136$\pm$6 $\mu$Jy) is consistent with that detected since 1989 
at similar frequencies. This result implies that the centimeter emission of
AB Aur is, within noise, steady in time.

\begin{figure}
\centering
\includegraphics[angle=0,scale=0.8]{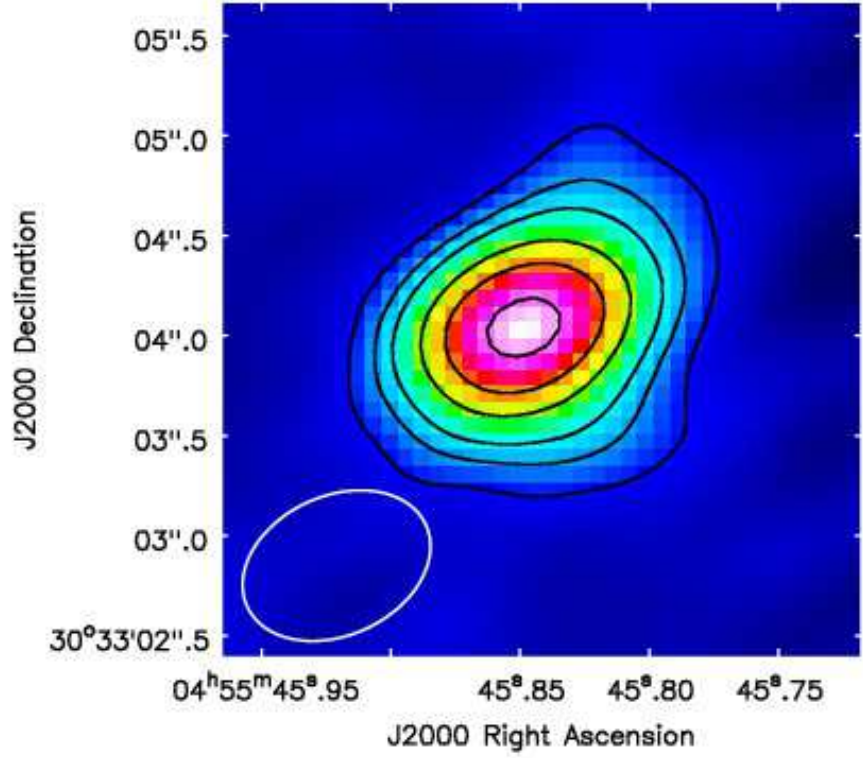}
\vskip-1.0cm
\caption{\small VLA image of the central radio source of AB Aur, made at 42 GHz
with natural weighting. Contours are -3, 3, 5, 7, 11, 15 and 21 times 25 $\mu$Jy beam$^{-1}$,
the rms noise of the image. The synthesized beam is shown in the bottom left
corner of the image.}
\label{fig2}
\end{figure}

The transitional disk of AB Aur has also been detected at millimeter wavelengths, with total flux densities of
350$\pm$20 mJy and 2200$\pm$100 mJy at 850 and 450 $\mu$m, respectively
(Sandell et al. 2011). Tang et al. (2012) report a total flux density of 110$\pm$1 mJy
at 1.3 mm (230 GHz). These large flux densities are dominated by the transitional disk
that is evident in Figure 1. However, Tang et al. (2012) note that within the cavity
they detect a compact source with a flux density of 1.3$\pm$0.2 mJy and suggest that this
emission could be due purely to free-free emission from an ionized jet. 
In Figure 3 we show the continuum spectrum of the emission closely associated with the star.
The points at 42.0 GHz and lower frequencies have been
least-squares fitted to a power law of the form $(S_\nu/\mu Jy) = 12\pm5 (\nu/GHz)^{1.1\pm0.1}$.
This spectral index in consistent with the values expected for a free-free jet
(Reynolds 1986). The flux density measured at 230 GHz by Tang et al. (2012) falls a factor
of $\sim$4 below the extrapolated flux density from the power law fit. This suggests that
the jet has become optically thin at $\sim$70 GHz. 

\begin{figure}
\centering
\includegraphics[angle=0,scale=0.8]{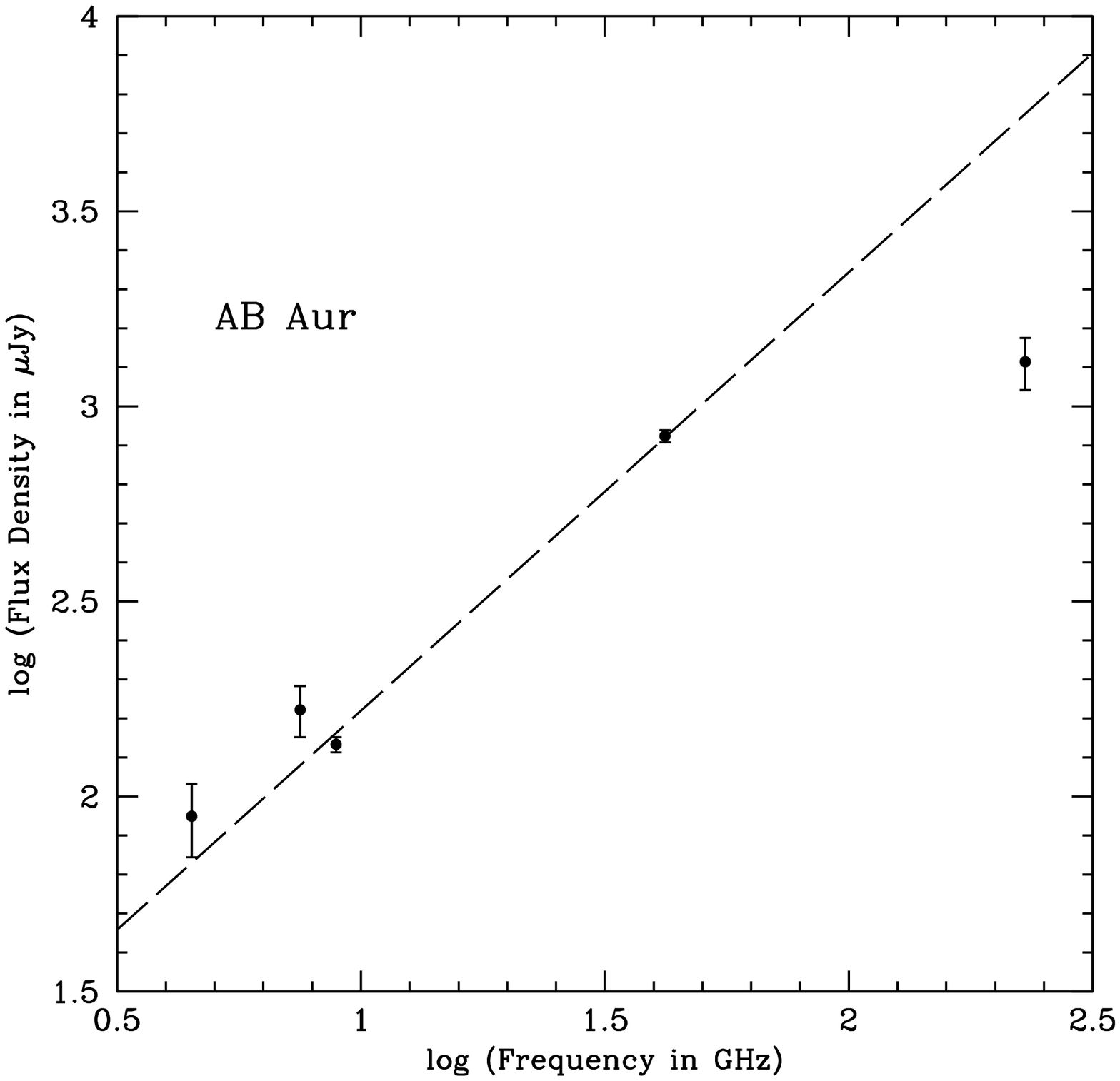}
\vskip-1.0cm
\caption{\small Continuum spectrum of the emission associated with
AB Aur. Data from Dzib et al. (2014), this paper, 
and Tang et al. (2012). The four points at lower frequencies have been
least-squares fitted to a power law of the form $(S_\nu/\mu Jy) = 12\pm5 (\nu/GHz)^{1.1\pm0.1}$. 
}
\label{fig3}
\end{figure}

An interesting alternative possibility to the jet is that of a photoevaporating
circumstellar disk (Pascucci et al. 2012). If the central star produces
sufficient EUV (13.6 eV $<$ h$\nu$ $\leq$ 100 eV) or X-ray
photons, the surface layers of the disk are ionized and produce
a slow ($\sim$10 km s$^{-1}$) photoevaporative wind that can be detected as a weak free-free source
(Pascucci et al. 2012; Owen et al. 2013). 
However, we favor the jet interpretation on two grounds. First, the region of free-free emission
in the photoevaporating disk is small ($\sim$10 AU; Pascucci et al. 2012; Owen et al. 2013) 
and will appear as unresolved
in our data. In contrast, the detected radio source is clearly resolved in one axis.
The second argument is that the free-free emission from photoevaporating disks is
expected to show a flat spectrum (Pascucci et al. 2012), while the detected source
has a spectrum that rises with frequency.
The lack of clear radio detections of photoevaporating disks (Galv\'an-Madrid et al. 2014;
Pascucci et al. 2014) has led these latter authors to propose that these winds could be largely neutral.

In summary, the radio continuum emission associated with AB Aur can be interpreted as a thermal jet
(e.g. Anglada 1996; Eisl\"offel et al. 2000). Its morphology, orientation with respect
to the inner CO disk, spectral index and lack of significant time variability
are all in agreement with this interpretation.

\subsection{Comparison with other jets}

Given that AB Aur is associated with a transitional disk, we expect its radio jet to be weaker than those
observed in forming stars with classical (full) disks. 
By a full disk we refer to a disk with no significant discontinuities in its radial dust distribution.
In sources with classical disks, the radio luminosity of 
the radio jet ($S_\nu d^2$, with $S_\nu$ the flux density at 8.3 GHz and $d$ the distance) 
is correlated with the bolometric luminosity of 
the source, $L_{\rm bol}$ by (Anglada 1995; Anglada et al. 2014):

\begin{equation} {\biggl(\frac{S_\nu d^2}{\rm mJy~kpc^2}\biggr)} = 0.008 
\biggl(\frac{L_{\rm bol}}{L_\odot}\biggr)^{0.6}. \end{equation}

Since for AB Aur $L_{bol}$ = 38 $L_\odot$ and $d$ = 144 pc, a flux density
of $\sim$3 mJy is expected at 8.3 GHz. In contrast, we detect a flux density
of $\sim$0.14 mJy, a factor of $\sim$20 below the value expected for a classical disk.

As noted above, accretion is also known to be present in transitional disks, although at rates an order
of magnitude smaller than those present in classical disks. Our results suggest that outflows are also
present in systems with transitional disks, but again at significantly lower rates than those
observed in classical disks. The correlation given above is most probably valid only
for very young stars, where most of the luminosity comes from accretion. In an object
like AB Aur most of the luminosity comes from the star.

It is relevant to compare the accretion and outflow rates in AB Aur.
From near-infrared hydrogen recombination lines,
its accretion rate is estimated to be $\dot M_{acc} \sim 1.4 \times 10^{-7}~M_\odot~yr^{-1}$
(Garcia Lopez et al. 2006; Salyk et al. 2013).
The radio luminosity is also correlated with $\dot P_{out}$, the momentum rate of the outflow,
by (Anglada 1995; Anglada et al. 2014):

\begin{equation} {\biggl(\frac{S_\nu d^2}{\rm mJy~kpc^2}\biggr)} = 190 
\biggl(\frac{\dot P_{\rm out}}{M_\odot yr^{-1}~km~s^{-1}}\biggr)^{0.9}. \end{equation}

Thus, from the observed radio luminosity, we expect $\dot P_{out} \simeq 5.0 \times 10^{-6}
M_\odot yr^{-1}~km~s^{-1}$.
Adopting
a jet terminal velocity of 300 km s$^{-1}$, we
estimate a mass loss rate of $\dot M_{out} \sim 1.7 \times 10^{-8}~M_\odot~yr^{-1}$.
The mass loss rate is then about a tenth of the accretion rate,
the approximate ratio deduced for full disks (Cabrit 2007). 
It is then interesting that although stars with transitional disks present lower
mass accretion and mass loss rates, the ratio $\dot M_{out}/\dot M_{acc}$ is $\sim$0.1,
as observed in less evolved objects.


\acknowledgements
We thank Y.-W. Tang for providing us with her 1.3 mm image of AB Aur.
LFR, LAZ, and LL acknowledge the financial support of CONACyT, M\'exico
and DGAPA, UNAM.
G.A. and E.M. acknowledge support from MICINN (Spain)
AYA2011-30228-C03-01 grant (co-funded with FEDER funds).

\end{document}